# Probing Tunneling Spin Injection into Graphene via Bias Dependence

Tiancong Zhu[1], Simranjeet Singh[1], Jyoti Katoch[1], Hua Wen[2], Kirill Belashchenko[3], Igor Žutić[4], Roland K. Kawakami[1,2]

[1]*Department of Physics, The Ohio State University, Columbus, OH 43210, USA*
[2]*Department of Physics and Astronomy, University of California, Riverside, CA 92521, USA*
[3]*Department of Physics and Astronomy and Nebraska Center for Materials and Nanoscience, University of Nebraska-Lincoln, Lincoln, NE 68588, USA*
[4]*Department of Physics, University at Buffalo, State University of New York, Buffalo, NY 14260, USA*

**Abstract**

The bias dependence of spin injection in graphene lateral spin valves is systematically studied to determine the factors affecting the tunneling spin injection efficiency. Three types of junctions are investigated, including MgO and hexagonal boron nitride (hBN) tunnel barriers and direct contacts. A DC bias current applied to the injector electrode induces a strong nonlinear bias dependence of the nonlocal spin signal for both MgO and hBN tunnel barriers. Furthermore, this signal reverses its sign at a negative DC bias for both kinds of tunnel barriers. The analysis of the bias dependence for injector electrodes with a wide range of contact resistances suggests that the sign reversal correlates with bias voltage rather than current. We consider different mechanisms for nonlinear bias dependence and conclude that the energy-dependent spin-polarized electronic structure of the ferromagnetic electrodes, rather than the electrical field-induced spin drift effect or spin filtering effect of the tunnel barrier, is the most likely explanation of the experimental observations.

## Introduction

Graphene has emerged as an ideal channel material for spintronic applications [1,2]. The long spin lifetime and spin diffusion length at room temperature make graphene one of the most efficient materials for transferring information with electron spins [3-5]. Furthermore, recent demonstrations of modulating the spin transport in graphene using magnetic proximity effect [6-8], gate-tunable spin absorption [9,10], and spin lifetime anisotropy [11-14] have generated new opportunities for future spintronic devices. These properties make graphene one of the most promising channel materials for developing next-generation spintronic devices [15-21].

The potential of graphene-based spintronic devices has also stimulated extensive studies of spin injection from ferromagnetic electrodes into graphene, which is critical for device operation. Since the first demonstration of electrical spin injection in graphene [3], much progress has been made in this direction. For example, the insertion of tunnel barriers between the ferromagnetic electrodes and graphene was found to minimize the conductance mismatch and enhance the spin lifetime and electrical spin injection efficiency defined as the spin polarization of the injected carriers [22]. Further development of the tunnel barrier material has increased the spin lifetimes, spin diffusion lengths, and spin accumulations achieved in spin transport measurements in graphene [5,23-28]. In addition, improved modeling of spin transport and spin precession that includes spin absorption effects at the contacts has enabled a more accurate determination of both spin lifetime and spin injection efficiency from the experimental data [29,30].

Despite these advances, the spin injection process in graphene lateral spin valves (LSV) is not fully understood, especially with respect to its dependence on the DC bias current. While the low-bias behavior of the lateral spin transport is well described by various equivalent resistor models [31-34], this treatment is restricted to the linear region in the bias dependence of the nonlocal spin signal (i.e., it assumes that the spin polarization of the electrodes and the spin diffusion lengths are independent of the bias). On the other hand, nonlinear dependence of the nonlocal spin signal on the DC current bias has been reported in several experiments [27,35-38]. Different mechanisms, including electric field-driven spin drift effect, spin filtering effect, and energy-dependent spin-polarized electronic structure have been proposed to explain the experimental results [39-42]. Because these models highlight different aspects of the spin injection process, understanding the nonlinear bias dependence is important for elucidating the factors that determine the spin injection efficiency. Interestingly, recent experiments on spin injection in Co/hBN/graphene junctions by Kamalakar *et al.* [37] and Gurram *et al.* [27] and in Co/MgO/graphene junctions by Ringer *et al.* [38] have independently reported a nonlinear bias dependence with a *sign reversal* at a negative DC bias. A systematic study of this sign-reversal feature and the conditions needed for it to appear across different tunnel barriers can help reveal the mechanism of tunneling spin injection in graphene-based LSVs.

In this work, we investigate the bias dependence of spin injection in graphene with different types of contacts to address this issue. We show that both MgO and hBN tunnel barriers exhibit similar nonlinear behavior in the bias-dependent spin injection measurement, including the sign reversal of the spin signal at a negative bias. By measuring multiple graphene LSVs with a wide range of contact resistances, we find that the bias-dependent behavior and the DC bias current at which the sign reversal occurs strongly depend on the resistance of the tunnel barrier. Further analysis shows that the sign reversal of the spin signal occurs only within a certain range of DC bias voltages, regardless of the tunnel barrier material or its resistance. These results suggest that the tunneling spin injection in graphene is likely determined by the energy-dependent spin-polarized electronic structure of the ferromagnetic electrode, rather than the electrical field induced spin-drift effect or spin filtering effect of the tunnel barrier.

**Experimental Details**

We fabricate graphene LSVs with transparent and tunnel barrier contacts to perform the bias-dependent spin injection study. Fig. 1 (a) shows a schematic diagram of such devices. Monolayer graphene is exfoliated from bulk crystals onto 300 nm $SiO_2$-Si substrate, and the electrodes are defined with standard e-beam lithography. The degenerately doped Si substrate is used as a back gate. For the transparent graphene LSVs with direct contact, we follow the fabrication procedure described in [43], with a 2 nm MgO masking layer between the Co electrode and graphene to reduce the direct contact area. For the MgO tunnel barrier devices, 0.8 nm to 1.2 nm of Ti seeded MgO is used for the tunnel barrier, followed by a 3 nm MgO masking layer. The fabrication details are the same as in [22]. For the hBN tunnel barrier devices, bilayer hBN is used following the fabrication procedure described in [26]. All the graphene LSVs are fabricated with two-step lithography, with Ti/Au electrodes at both ends of the device. This avoids spin signal contribution from the outer electrodes during measurement.

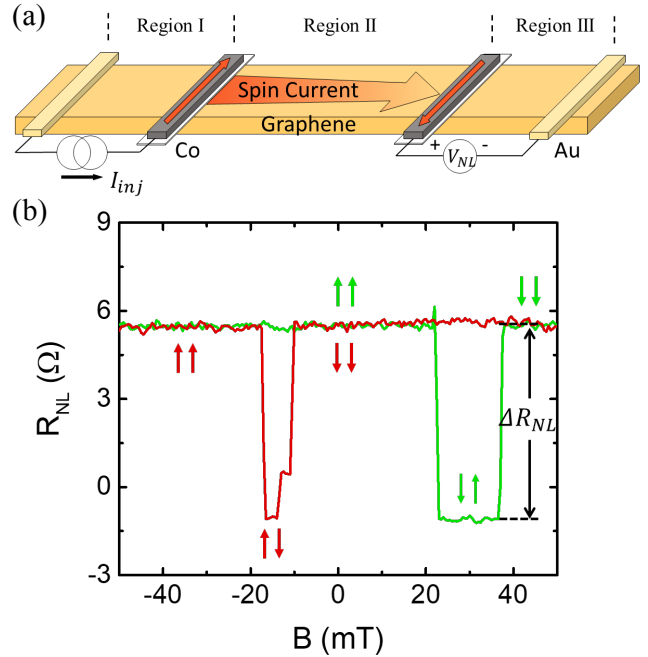

Fig. 1. (a) Schematic diagram of a graphene LSV and the nonlocal measurement geometry. An AC+DC current is applied on the left Co electrode to perform the bias-dependent spin injection study. The region I, II, and III are corresponding to the spin injection, diffusion, and detection channel. (b) Typical nonlocal measurement data on a graphene lateral spin valve.

We use low-frequency (11 Hz) lock-in techniques to perform electrical and spin transport measurement on the graphene LSVs. The rms for the injection AC current is $1\,\mu A$. First, the channel resistance and contact resistance are characterized with the standard four-probe and three-probe geometry. Spin transport in graphene is then measured in the nonlocal geometry, as shown in Fig. 1 (a). In the nonlocal geometry, an AC charge current ($I_{inj}$) is applied in the left circuit, and the AC nonlocal voltage ($V_{NL}$) is measured in the right circuit. The ratio $R_{NL} = V_{NL}/I_{inj}$ is defined as the nonlocal resistance. Fig. 1 (b) shows the typical nonlocal resistance data obtained from the measurement. During the measurement, an external magnetic field is swept parallel to the ferromagnetic electrodes, which changes the relative alignment direction of the electrode magnetization. Two different levels of nonlocal resistance can be obtained, depending on whether the magnetization of the injector and detector electrodes are parallel ($\uparrow\uparrow, \downarrow\downarrow$) or anti-parallel ($\uparrow\downarrow, \downarrow\uparrow$) to each other. The nonlocal magnetoresistance ($\Delta R_{NL}$) is defined as the difference of $R_{NL}$ between the parallel and anti-parallel state, $\Delta R_{NL} = R_{NL}^{\uparrow\uparrow} - R_{NL}^{\uparrow\downarrow}$.

## Results

To perform the bias-dependent spin injection study, a DC current bias is applied on the injector electrode in addition to the AC current, with positive bias defined as current flowing from the Co electrode into graphene. The lock-in detection measures the AC response in $V_{NL}$. Figure 2(a) shows the nonlocal magnetoresistance curves measured for different DC bias currents on a MgO tunnel barrier graphene LSV with injector contact resistance $R_C = 63\,k\Omega$. The gate voltage is zero. Interestingly, the nonlocal magnetoresistance signal shows a strong variation when the DC bias current is changed. At zero DC bias current, the $\Delta R_{NL}$ is $6\,\Omega$. When a positive DC bias current is applied across the injector electrode, the magnitude of $\Delta R_{NL}$ first increases up to $10.6\,\Omega$ ($I_{DC} = 4\,\mu A$), then slowly decreases down to $5.1\,\Omega$ ($I_{DC} = 16\,\mu A$). The spin signal at $I_{DC} = 4\,\mu A$ exhibits a 77% increase in the signal magnitude compared to zero bias. Notably, when a negative DC bias current is applied, the nonlocal magnetoresistance curve inverts for $I_{DC}$ more negative than $-4\,\mu A$. The inverted curve indicates an opposite orientation of spin polarization of the injected carriers, which is represented by a negative value for $\Delta R_{NL}$. At $I_{DC} = -12\,\mu A$, the $\Delta R_{NL}$ reaches $-2.8\,\Omega$, which is $-47\%$ of the zero-bias signal.

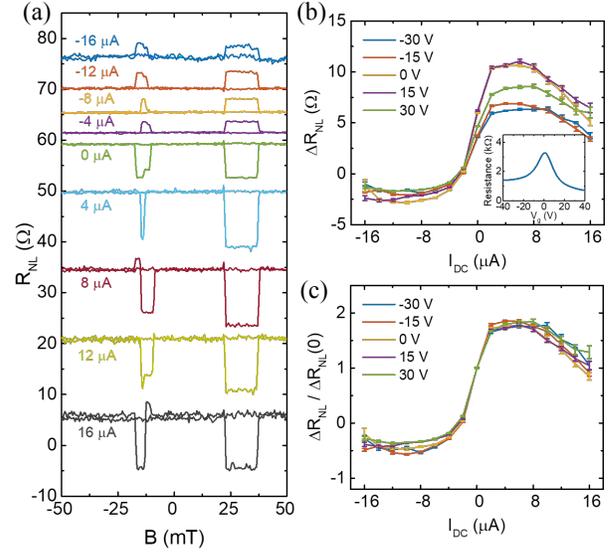

Fig. 2. (a) Nonlocal resistance of a graphene LSV with MgO tunnel barrier measured at different DC bias current. The back-gate voltage $V_g = 0V$. All curves are shifted for clarity. (b) Bias-dependent $\Delta R_{NL}$ measured at different gate voltages. The charge neutrality point is at $V_{CNP} = 0$ V. Inset: gate dependent resistance of the graphene channel. (c) Bias-dependent $\Delta R_{NL}$ (data from panel b) with each $\Delta R_{NL}(I_{DC})$ curve normalized by its zero-bias value.

To investigate the bias dependence of $\Delta R_{NL}$ at different carrier densities for the spin diffusion channel, we perform the same measurement at different gate voltages, $V_g$. We define the charge neutrality point voltage $V_{CNP}$ as the gate voltage with maximum resistance in the graphene channel, and the carriers in the graphene are dominated by electrons when $V_g > V_{CNP}$, and dominated by holes for $V_g < V_{CNP}$. Fig. 2(b) shows the result of the measurement, with each curve illustrating the bias dependence of $\Delta R_{NL}$ for a different gate voltage. Each of the curves exhibit a strong nonlinear bias dependence and the curve shapes are similar for all different gate voltages. Only the overall magnitude of the curves shows a variation with gate voltage, which could be due to a change of spin lifetime and spin diffusion length as a function of carrier density in the graphene spin transport channel. In order to better compare the bias-dependent spin injection at different gate voltages, we plot the same data in Fig. 2(b) by normalizing each curve by its zero-bias value, $\Delta R_{NL}(0)$. Figure 2(c) shows the normalized data. After the normalization, the bias-dependent $\Delta R_{NL}$ curves almost collapse onto one single curve, independent of gate voltage. This shows that the observed modulation of $\Delta R_{NL}$ with DC bias current does not depend on the carrier density or carrier type in the spin diffusion channel of graphene. On the other hand, this behavior is consistent with mechanisms that alter the effective spin polarizations of the injector contact as a function of bias.

We also perform a similar study on two graphene LSVs with hBN tunnel barriers. Fig. 3 shows the bias-dependent $\Delta R_{NL}$ and the normalized data from one of the devices. For this device, the contact resistance of the injector electrode is $7\ k\Omega$. The nonlocal magnetoresistance is increased by more than 100% at positive bias, and reverses sign at negative bias. The line-shape of the bias-dependent $\Delta R_{NL}$ for the hBN tunnel barrier device is similar to that of the MgO tunnel barrier in Fig. 2(b).

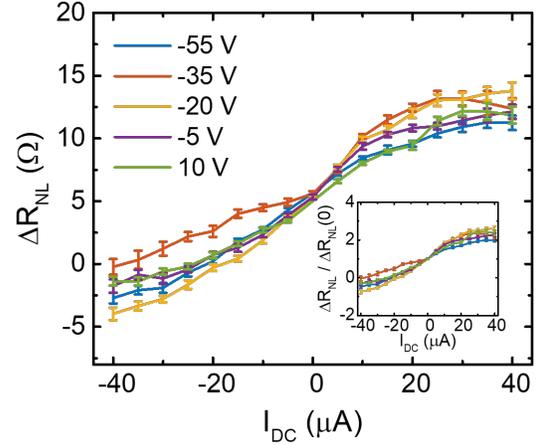

Fig. 3. Bias-dependent $\Delta R_{NL}$ measured on a graphene LSV with hBN tunnel barrier at different back-gate voltage. The inset shows the normalized data. $V_{CNP}$ = -27 V for this device.

Comparing Fig. 2(b) and Fig. 3, we notice that although the bias-dependent $\Delta R_{NL}$ shapes are similar, the DC bias current required for the signal sign-reversal ($I_{rev}$) are very different ($\approx -20\ \mu A$ for hBN in Fig. 3 and ≈ −2 $\mu A$ for MgO in Fig. 2). Such differences can be due to either having different tunnel barrier materials or having different contact resistances of the injector electrodes. In order to address this issue, we measure the bias dependence of spin injection on 8 additional graphene LSVs (2 with direct contact, and 6 with MgO tunnel barriers) of different contact resistances, ranging from 0.18 $k\Omega$ to 131 $k\Omega$. Figure 4 shows some of the representative results. For graphene LSVs with direct contact ($R_c = 0.18\ k\Omega$, Fig. 4(a)), the $\Delta R_{NL}$ is almost constant within a large DC bias current range of $[-100\ \mu A, 100\ \mu A]$. For MgO tunnel barriers with low contact resistance ($R_c = 5.3\ k\Omega$), Fig. 4(b) shows that the bias dependence of $\Delta R_{NL}$ is nonlinear and switches sign at $I_{rev} \cong -25\ \mu A$. For MgO tunnel barriers with high contact resistance ($R_c = 131\ k\Omega$) Fig. 4(c) shows the nonlinear behavior of bias-dependent $\Delta R_{NL}$ in an even smaller DC bias range ($[-5\ \mu A, 5\ \mu A]$). The $\Delta R_{NL}$ also switches sign at lower value $I_{rev} \cong -0.75\ \mu A$, which is more than an order of magnitude smaller than that in Fig. 4(b). However, a much smaller difference is observed when considering DC bias voltage ($V_{DC}$) instead of DC bias current on the injector contact, as shown in the top axes of Fig. 4. In this case, the DC bias voltage for the low contact resistance MgO tunnel barrier device to reverse sign is $V_{rev} \cong -112\ mV$, which is much closer to that of the high contact resistance device ($V_{rev} \cong -98\ mV$). This behavior indicates that the nonlinear bias-dependent $\Delta R_{NL}$ is strongly correlated to the DC bias voltage on the contact electrodes.

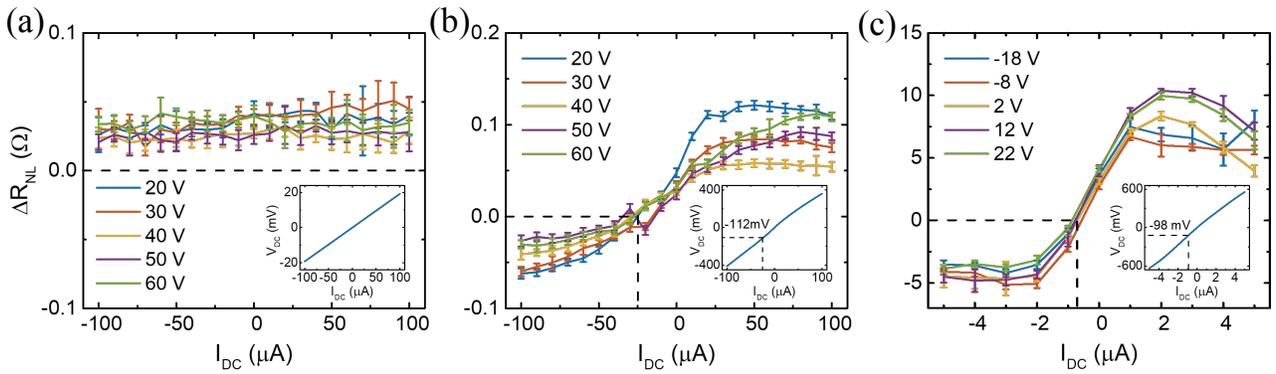

Fig. 4. Bias dependence of $\Delta R_{NL}$ for graphene lateral spin valves with (a) transparent contacts ($V_{CNP}$ = 44 V), (b) MgO tunnel barrier contacts with low contact resistance ($V_{CNP}$ = 45 V), and (c) MgO tunnel barrier contact with high contact resistance ($V_{CNP}$ = 2 V). The inset shows the IV curves integrated from the 3-probe dV/dI measurement. The dashed lines in the inset shows the position of $I_{rev}$ and $V_{rev}$ for the corresponding contacts.

To examine if the correlation between the DC bias voltage and the sign-reversal applies to other devices measured in our study, we plot the $I_{rev}$ and $V_{rev}$ as a function of contact resistance for the measured MgO and hBN tunnel barrier devices (transparent contact devices are not included because we do not observe a sign reversal of the signal). As shown in Fig. 5(a), a strong variation of $I_{rev}$ with different contact resistances can be observed, which is inversely proportional to $R_c$ (dashed line). In addition, Fig. 5(b) shows that within the large range of measured $R_c$, the values of $V_{rev}$ always occur in a small voltage window ($[-225\ mV, -75\ mV]$) for both MgO and hBN devices. This establishes the correlation between the DC bias voltage and the sign reversal of the nonlocal signal, and also suggests that the sign reversal does not depend on the tunnel barrier material.

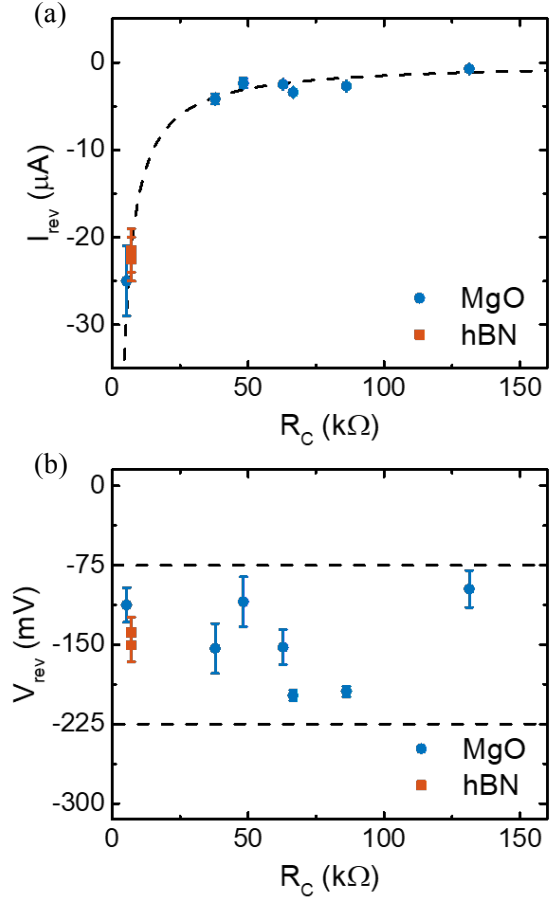

Fig. 5. (a) The reversal bias current ($I_{rev}$) plotted as function of contact resistance. Each data point represents a different injector electrode. The error bars represent the variation of $I_{rev}$ with back-gate voltage. The dashed line plotted as $I = -\frac{150\ mV}{R}$ is a guide to the eye. (b) The reversal bias voltage ($V_{rev}$) plotted as function of contact resistance. All the reversal bias voltages are within the range between $-75\ mV$ and $-225\ mV$, as indicated by the dashed lines.

## Discussion

We now discuss several mechanisms that can give rise to the nonlinear bias-dependent spin signal, including the electric field-induced spin drift, the tunnel barrier spin filtering, and the spin-polarized electronic structure of the ferromagnetic electrodes.

We first consider the electric field-induced spin drift effect [44-47], which emphasizes the impact of the spin transport channel on spin injection efficiency. In graphene LSVs, it has been shown that an electric field in the spin diffusion channel (Region II in Figure 1(a)) can produce a drift effect of the spin-polarized charge carriers, which affects the measured nonlocal signal [48,49]. The increase (decrease) of the nonlocal signal depends on the carrier type and the direction of the electric field. Similarly, the electric field associated with the DC bias current in the spin injection circuit [Region I in Fig. 1(a)] can also modify spin transport in graphene, which could lead to a nonlinear bias dependence of the nonlocal spin signal. This effect was proposed by Józsa et al. [36] and Yu et al. [39] to explain the strong nonlinear bias-dependent spin signal observed in graphene LSVs.

To investigate the effect of spin drift on spin injection, we develop a drift-diffusion model following [34] to describe spin transport in graphene LSVs. In the presence of electric field, the spin-dependent electrochemical potential in graphene can be written as

$$u_s(x) = A e^{x/\lambda_+} + B e^{-x/\lambda_-} \quad (1)$$

where $\lambda_\pm = \lambda \left( +(-) \frac{\mu \lambda E}{2D} \pm \sqrt{\left(\frac{\mu \lambda E}{2D}\right)^2 + 1} \right)^{-1}$ are the spin transport lengths for the upstream and downstream carriers, $\mu$ is the mobility and $\lambda$ the spin diffusion length in graphene without the electric field, $E$ is the electric field induced by the injection current, and $D$ is the diffusion coefficient. The plus or minus sign in the term outside of the square root is for electron or hole-dominated transport, respectively. The electric field in the spin injection circuit (region I) is proportional to the injection current, $E = \frac{R_{sq}}{w} I$, where $R_{sq}$ is the sheet resistance and $w$ the width of the graphene channel. In regions II and III there is no electric field, and $\lambda_+ = \lambda_- = \lambda$. By imposing the continuity condition on the spin current and spin-dependent chemical potential at the interfaces between different regions, we find the nonlocal voltage for the DC measurement

$$\Delta V_{NL} = \frac{4P^2 D}{\mu} \frac{\varepsilon}{(1+(-)\varepsilon + \sqrt{\varepsilon^2+1})} e^{-x/\lambda} \quad (2)$$

where $\varepsilon = \frac{\mu \lambda R_{sq}}{2Dw} I$, and $I$ is the total charge current. For an AC+DC measurement, the nonlocal signal from the lock-in yields the differential (AC) response $\Delta R_{NL} = d\Delta V_{NL}/dI$, which is shown in Fig. 6 as a function of the DC current $I_{DC}$. The curves were obtained using typical parameters for graphene LSVs in our measurements. This calculation shows that the electric field-induced spin drift in the spin injection channel can lead to a nonlinear dependence of the nonlocal resistance on the DC bias.

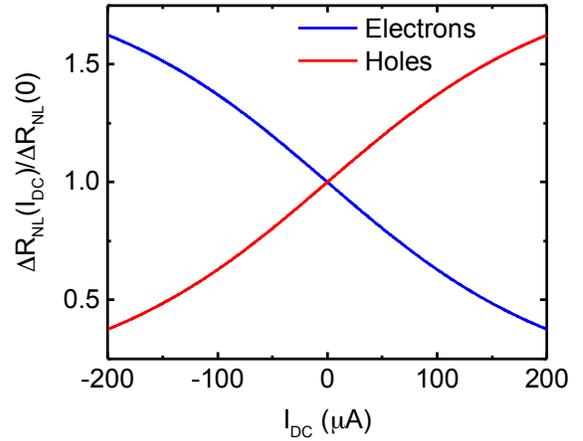

Fig. 6. Bias-dependent $\Delta R_{NL}(I_{DC})$ simulated with device parameters from typical graphene LSVs in this work, with $\mu = 4000 \, cm^2/Vs$, $\lambda = 4 \, \mu m$, $R_{sq} = 1 \, k\Omega$, $D = 0.02 \, m^2/s$, and $w = 1 \, \mu m$. The simulations are performed for both electron and hole dominated channel.

However, the electric field-induced spin drift is unlikely to be the dominant factor that determines the nonlinear spin signal in our measured devices. Strong spin drift effect requires that the drift velocity is comparable to the Fermi velocity. This requires either a high-mobility sample with hBN encapsulation [49] or a strong electric field that is only likely to exist in the immediate vicinity of a nanometer-sized pinhole [36]. In our graphene LSVs, the mobility is less than $5000 \, cm^2/V \cdot s$ limited by the $SiO_2$ substrate, while both the exfoliated hBN and MBE-grown MgO tunnel barriers are pinhole-free [50]. The drift velocities ($v_D = \mu E = \frac{\mu R_{sq}}{w} I$) in our pinhole-free devices are at least two orders of magnitude smaller compared to the Fermi velocity in graphene. This is also reflected as the large DC bias current range in Fig. 6 compared to our experimental data. Furthermore, in contrast with the predictions of the spin-drift model shown in Fig. 6, our experiments reveal no influence of the carrier type on the nonlinear bias-dependent signal. Therefore, this mechanism is unlikely to be responsible for the observed large variation of the nonlocal resistance.

The mechanism of the spin filtering effect emphasizes the impact of the non-magnetic tunnel barrier material on spin injection efficiency. Experimentally, Kamalakar *et al*. [37] have reported spin signal inversion and nonlinear bias dependence in graphene LSVs with a high resistance hBN tunnel barrier. They attribute the phenomenon to a spin filtering effect with the hBN tunnel barrier. However, such effect is material-specific, with a given tunnel barrier material favoring specific electronic states in the ferromagnetic electrode in the tunneling process. A well-known example is the symmetry filtering effect in Fe/MgO/Fe magnetic tunnel junctions [51,52], where the MgO barrier strongly favors the states of the $\Delta_1$ symmetry at the Γ point. On the other hand, the calculations in [53] show that the hBN tunnel barrier does not strongly filter the electronic states by their wavevector.

The symmetry filtering mechanism requires good crystallinity of the ferromagnetic electrode and the tunnel barrier and is also expected to be much stronger for MgO compared to hBN. In our graphene LSVs, the hBN layers are single-crystalline, but the Co electrodes and the MgO tunnel barriers are not. The Co/MgO/graphene junctions do not seem to meet the requirement for symmetry filtering yet still exhibit nonlinear bias dependence and sign reversal. Furthermore, the bias-dependent spin injection behavior of both MgO and hBN tunnel barriers look similar, with the nonlocal signal reversing its sign at roughly the same bias voltage. These observations suggest that the symmetry-based spin filtering effect is not a key factor for the bias dependence and sign reversal of the spin signal as observed in our experiment. Arguments against the spin filtering were also given in the analysis of Co/graphene/hBN/NiFe vertical spin valves where the bias-dependent magnetoresistance could show a sign reversal [54].

The fact that the sign reversal of the nonlocal signal occurs at similar bias voltages for different tunnel barriers [see Fig. 5(b)] suggests that the nonlinear dependence originates from the energy-dependent spin-polarized electronic structure of the Co electrode. Under this mechanism, the spin injection efficiency is determined by the band alignment between Co and graphene, which is controlled by the voltage drop across the tunnel barrier. Furthermore, this mechanism produces a nonlinear bias dependence of the spin signal without any special requirements on the tunnel barrier. Thus, similar bias-dependent spin signals for hBN and MgO tunnel barriers (Fig. 2 and 3) with similar sign-reversal voltages [Fig. 5(b)] as observed in our experiments are expected if the energy-dependent spin-polarized electronic structure of the ferromagnetic electrodes is the dominant factor. Because our main experimental results (i.e., the similarity of the sign-reversal voltage for both tunnel barriers of varying resistances) are readily understood within this framework, we believe the spin-polarized electronic structure of the Co electrode is the main factor that determines the nonlinear bias dependence and sign reversal in the spin signal.

There are several aspects of the energy-dependent spin-polarized electronic structure that affect the tunneling spin injection from Co into graphene. First, it could be related to the band structure of bulk Co. Although the spin polarization derived from the density of states of bulk Co does not reverse its sign in a wide energy range around the Fermi level, it was argued that one should consider the spin-polarized density of state (DOS) convoluted with the electron velocity $v^\alpha$, where α = 1 for ballistic transport or α = 2 for diffusive transport, to calculate the spin injection efficiency [55]. Sipahi *et al*. [42] have considered the case for Co in direct contact to graphene, where the calculated spin injection efficiency does show strong energy dependence when considering the electron velocity at different energy levels. Furthermore, surface states at the Co/MgO or Co/hBN interfaces can also play an important role in determining the spin injection efficiency [56]. Using Fe/GaAs(001) as an example, it was shown that the spin polarization of the tunneling current can exhibit a nonlinear bias dependence and change sign under a relatively small bias voltage due to the minority-spin resonant state at the Fe/GaAs(001) interface [57]. Such behavior has been observed experimentally in nonlocal spin transport by Lou *et al.* [58]. A similar scenario could also happen in our devices. These possibilities are strongly dependent on the crystallographic orientation of the ferromagnetic electrode. The polycrystalline nature of the Co electrodes in our devices makes it difficult to compare the experimental result with the mechanisms discussed above. Experimentally, this difficulty could be overcome by synthesizing single-crystal Co electrodes on MgO substrate [59] and fabricating graphene LSVs with the inverted structure, as developed by Drögeler *et al.* [60].

## Conclusion

In summary, we have performed a systematic study on bias-dependent spin injection into graphene with both MgO and hBN tunnel barriers. We observe a strong nonlinear behavior of the spin signal with sign-reversal in both systems. By normalizing the bias-dependent spin signal with its zero-bias value, we find that the relative change in spin injection efficiency does not depend on the carrier density inside the graphene channel, indicating that our observation is related to the junction region of the ferromagnetic electrode. By comparing bias-dependent spin injection measurements on multiple devices, we find that the sign-reversal of the spin signal is associated with a certain bias voltage window, independent of the contact resistance and tunnel barrier material. By comparing different mechanisms with our experimental data, we conclude that the bias dependence of the tunneling spin injection in graphene is most likely induced by the energy dependence of the spin-dependent electronic structure of the ferromagnetic electrode. While the observed nonlinear response complicates the description of the graphene-based lateral spin valves, beyond the usual equivalent-resistor models, it also provides important device opportunities for spin logic [19] and spin communication [61,62] with bias-dependent modulation of spin polarization.

## Acknowledgement

Funding for this research was provided by the US Department of Energy (Grant No. DE-SC0018172). S.S. and J.K. acknowledge the support from the Center for Emergent Materials: an NSF MRSEC under award number DMR-1420451. K.B. acknowledges support from the National Science Foundation through the Nebraska MRSEC (Grant No. DMR-1420645) and Grant No. DMR-1609776. I. Ž. acknowledges support from US Office of Naval Research (Grant No. 000141712793) and National Science Foundation (Grant No. ECCS-1508873 and ECCS-1810266).